\documentclass[conference]{IEEEtran}
\IEEEoverridecommandlockouts
\usepackage{cite}
\usepackage{amsmath,amssymb,amsfonts}
\usepackage{algorithmic}
\usepackage{graphicx}
\usepackage{textcomp}
\usepackage{xcolor}
\usepackage{booktabs} 
\usepackage{listings}
\usepackage{soul}
\usepackage[bookmarks=false]{hyperref}
\usepackage{xspace}
\usepackage{graphicx}
\usepackage{ifthen}
\usepackage[normalem]{ulem} 
\usepackage{xcolor,colortbl}
\usepackage{hyperref}

\newboolean{showedits}
\setboolean{showedits}{true} 
\ifthenelse{\boolean{showedits}}
{
	\newcommand{\del}[1]{\textcolor{red}{\sout{#1}}} 
	\newcommand{\nbe}[3]{
		{\colorbox{#3}{\bfseries\sffamily\scriptsize\textcolor{white}{#1}}}
		{\textcolor{#3}{\sf\small$\blacktriangleright$\textit{#2}$\blacktriangleleft$}}}
}{
	\newcommand{\del}[1]{} 
	
	\newcommand{\nbe}[3]{}
}

\newboolean{showcomments}
\setboolean{showcomments}{true} 
\newcommand{\id}[1]{$-$Id: scgPaper.tex 32478 2010-04-29 09:11:32Z oscar $-$}

\ifthenelse{\boolean{showcomments}}
 {
 	\newcommand{\nbc}[3]{
 		{\colorbox{#3}{\bfseries\sffamily\scriptsize\textcolor{white}{#1}}}
		{\textcolor{#3}{\sf\small$\blacktriangleright$\textit{#2}$\blacktriangleleft$}}}
	
 }{
 	\newcommand{\nbc}[3]{}
 	
 }

\usepackage[most]{tcolorbox}
\ifthenelse{\boolean{showedits}}
{
  \newtcolorbox{inserted}{%
       title=Inserted text:,
       colframe=blue,colback=blue!5!white,
       breakable,
       leftrule=0mm, 
       bottomrule=0mm,
       rightrule=0mm,
       toprule=0mm,
       arc=0mm, outer arc=0mm,
       oversize
  }
  \newtcolorbox{deleted}{%
       title=Deleted text:,
       colframe=red,colback=red!5!white,
       breakable,
       leftrule=0mm, 
       bottomrule=0mm,
       rightrule=0mm,
       toprule=0mm,
       arc=0mm, outer arc=0mm,
       oversize
  }
  \newtcolorbox{refactored}{%
       title=Rewritten text:,
       colframe=blue,colback=red!5!white,
       breakable,
       leftrule=0mm, 
       bottomrule=0mm,
       rightrule=0mm,
       toprule=0mm,
       arc=0mm, outer arc=0mm,
       oversize
  }
}{

}

\newcommand{\commented}[1]{}

\newcommand{\ie}{\emph{i.e.,}\xspace}
\newcommand{\etal}{\emph{et al.}\xspace}

\definecolor{source}{gray}{0.95}
\definecolor{highlight}{gray}{0.9}

\newcommand{\code}[1]{\texttt{#1}}

\usepackage[T1]{fontenc} 

\lstnewenvironment{codesnippet}{%
	\lstset{%
		frame=single,
		framerule=0pt,
		mathescape=false
	}
}{}

\newcommand{\CG}{CogniCrypt\xspace}
\newcommand{\GH}{GitHub\xspace}

\newcommand{\vs}{vs.\xspace}

\newcommand{\boxit}[1]{\vspace{0.3cm}
\noindent
\fbox{
\begin{minipage}{24em}
\emph{#1} 
\end{minipage}
}
}
\IEEEoverridecommandlockouts
\IEEEpubid{\makebox[\columnwidth]{\textbf{Preprint -- ESEM}\hfill} \hspace{\columnsep}\makebox[\columnwidth]{ }}

\makeatletter                  
\def\mdseries@tt{m}      
\makeatother                   

\begin{document}
\title{The Impact of Developer Experience\\in Using Java Cryptography}

\author{\IEEEauthorblockN{Mohammadreza Hazhirpasand}
\IEEEauthorblockA{\textit{SCG, University of Bern} \\
Bern, Switzerland \\
mohammadreza.hazhirpasand@inf.unibe.ch}
\and
\IEEEauthorblockN{Mohammad Ghafari}
\IEEEauthorblockA{\textit{SCG, University of Bern} \\
Bern, Switzerland \\
mohammad.ghafari@inf.unibe.ch}
\and
\IEEEauthorblockN{Stefan Kr\"uger}
\IEEEauthorblockA{\textit{University of Paderborn} \\
Paderborn, Germany \\
stefan.krueger@uni-paderborn.de}
\and
\IEEEauthorblockN{Eric Bodden}
\IEEEauthorblockA{\textit{University of Paderborn} \\
Paderborn, Germany \\
eric.bodden@uni-paderborn.de}
\and
\IEEEauthorblockN{Oscar Nierstrasz}
\IEEEauthorblockA{\textit{SCG, University of Bern} \\
Bern, Switzerland \\
oscar.nierstrasz@inf.unibe.ch}
}

\maketitle

\begin{abstract}
Background: 
Previous research has shown that crypto APIs are hard for developers to understand and difficult for them to use.
They consequently rely on unvalidated boilerplate code from online resources where security vulnerabilities are common.

Aims and method: 
We analyzed 2,324 open-source Java projects that rely on Java Cryptography Architecture (JCA) to understand how crypto APIs are used in practice, and what factors account for the performance of developers in using these APIs.

Results: 
We found that, in general, the experience of developers in using JCA does not correlate with their performance.
In particular, none of the factors such as the number or frequency of committed lines of code, the number of JCA APIs developers use, or the number of projects they are involved in correlate with developer performance in this domain.

Conclusions: 
We call for qualitative studies to shed light on the reasons underlying the success of developers who are expert in using cryptography.
Also, detailed investigation at API level is necessary to further clarify a developer obstacles in this domain.

\end{abstract}


\maketitle

\section{Introduction}
\label{sec:intro}


Cryptography has perhaps never been more important as a means to secure data, but previous research has found that cryptographic application programming interfaces (API) are hard to understand and, consequently, are often misused by developers~\cite{Egele, KrugerS0BM18, Rahaman2018chiron, ShaoDGYS14, Chatzikonstantinou:2016, LazarCWZ14}.

Egele \etal examined the use of block ciphers and message authentication codes in Android applications and found more than 10,000  applications misused cryptographic primitives in insecure manners~\cite{Egele}.
%
Nadi \etal examined the reasons for misuse of these APIs by surveying developers as well as investigating questions on StackOverflow~\cite{Nadi2016}.
They concluded that Java cryptography APIs are too low-level, and that developers would welcome crypto assistants as well as example-driven documentation.
Acar \etal~\cite{Yasemin2017} suggest the usability of cryptographic APIs has an important role in making software more secure, but have to conclude stark usability shortcomings in existing crypto APIs.
They identify simplicity of APIs, clarity of documentation, default secure options, and secure source code examples as factors that could sigificantly ease the usage of cryptographic APIs.

Misuse of cryptographic APIs is widespread and the reasons are manifold.
Existing studies agree that only a small portion of usages is actually correct and secure.
We hypothesize that \emph{identifying a portion of developers who properly use crypto APIs, and studying their characteristics may help us to identify the community of developers with cryptography expertise}.
However, to the best of our knowledge, no such investigation has been conducted.

In this paper, we study how developers perform in using cryptographic APIs, and investigate whether developer experience impacts whether they use an API correctly or not.
To this end, we mined 2,324 Java projects, and checked them for misuses of the Java Cryptography Architecture (JCA).
We discovered that, on average, of 3.9 crypto uses in each project, 2.5 are not secure, and that developers have considerable difficulties in using more than half of the APIs.
We also performed statistical analyses on the results to check for correlations between several measures of developer experience and success in using the APIs.
We found that factors such as the number and frequency of committed lines of code, API diversity, and the number of projects developers are involved in, do \emph{not} influence their performance.

We plan to conduct qualitative studies to shed lights on reasons underlying the success of developers who are expert in using cryptography.

The remainder of this paper is structured as follows.
In \autoref{sec:study}, we explain the setup of our study,
and present the results in \autoref{sec:results}.
In \autoref{sec:related} we discuss related work, and in \autoref{sec:plan} we outline future directions of this research.
Finally, we conclude the paper in \autoref{sec:conclusion}.

\section{Empirical Study}
\label{sec:study}
This section summarizes the setup of the experiment, and introduces some terminology that we use in the paper.


\subsection{Setup}

We mined Java projects on \GH, and downloaded and compiled those that use JCA.
We then ran the \CG tool on these projects to identify crypto issues.
The experiment entailed the following steps:

\subsubsection{Identifying relevant projects}
We initially started with 183 projects using JCA APIs that were identified in previous work~\cite{KrugerS0BM18}.
We were interested in collecting as many JCA projects as possible that we could associate to each developer.
Therefore, instead of an exhaustive search on \GH, we checked what other projects each developer had contributed to, and whether they use JCA.
We continued this process for every new project, and stopped when we obtained 2,324 projects that use JCA APIs.
We used the \GH API search to check whether a project uses any of the crypto classes specified in the \CG rule set.
Based mainly on the import statements in each project, we were able to identify and download 2,780 projects.
In case a project was forked, we looked for the original repository to download.

\subsubsection{Building projects}
We needed to compile each project in order to run a static analysis that identifies crypto misuses.
We wrote a bash script to build all the downloaded projects.
The bash script first checked the existence of the POM file in the project's path, and if it existed, we would continue with compilation, otherwise we excluded the project.
We tried to compile projects using the Maven build tool, while skipping the running of the tests (to save time).
Many projects could not be compiled due to dependencies that were not resolved.
We excluded such projects and did not manually investigate the issue.
Ultimately, we were able to compile and build 2,360 projects.

\subsubsection{Analyzing projects}
We used a static analysis tool called \CG to detect known misuses of cryptographic APIs in Java bytecode~\cite{KrugerS0BM18}.
This tool takes a target program and evaluates the program's correctness with respect to predefined method-call patterns, parameter constraints, and secure compositions of cryptography-related classes encoded as a set of rules.
We chose \CG for two main reasons: first, we had contact with the main developers of this tool, who were willing to support us in the execution of the experiment.
Second, \CG covers a wide range of crypto APIs while keeping false positives at a manageably low rate (usually below 10\%).

In this step, a bash script went through the folders of all projects and recursively looked for the \textit{target/classes} folder where the \textit{.class} files reside, to feed them into \CG.\footnote{We used the publicly available command-line version of \CG and its rule set available at the time of running this analysis on Jan 16th, 2019.} 
We specified a timeout of 15 minutes to abort lengthy analyses.
In the end, roughly 15\% of the analyses were interrupted, and we succeeded in analyzing 2,141 projects.

\subsubsection{Reporting the results}
A bash script parsed the analysis reports in order to collect the assessment of each project.
In particular, for every project, it checked which crypto APIs were (mis)used, and where in the source code such API usages are located, \ie the file name and the line number.
With the help of the \textit{git blame} command we also identified the last developer who committed the code associated with each API use, as well as the commit time.
We stored this information in a database.


\subsection{Terminology}
In this paper, when we refer to \emph{APIs} we mean those provided by the JCA framework, and a \emph{developer} is one who commits such API usages to a source file that belongs to one of the included projects in this study.
A commit/file/project is \emph{secure} when there is no API misuse, otherwise it is \emph{buggy}.
Finally, we use the term \emph{commits} to refer to \emph{the last crypto-related} committed lines of code.



\section{Results}
\label{sec:results}

We first present the state of cryptography uses in open-source Java projects, and then present our study of the factors that may influence developers' performance.

\subsection{The State of Cryptography Uses}

We analyzed a total of 2,324 projects, \ie 2,141 plus the 183 initial projects.
These projects consisted of a total of 2,652 unique files containing JCA APIs, which means, on average, 1.4 files per project with a standard deviation of 1.4.

\begin{figure}
\centering
\includegraphics[width=1\linewidth,trim=4 4 4 4,clip]{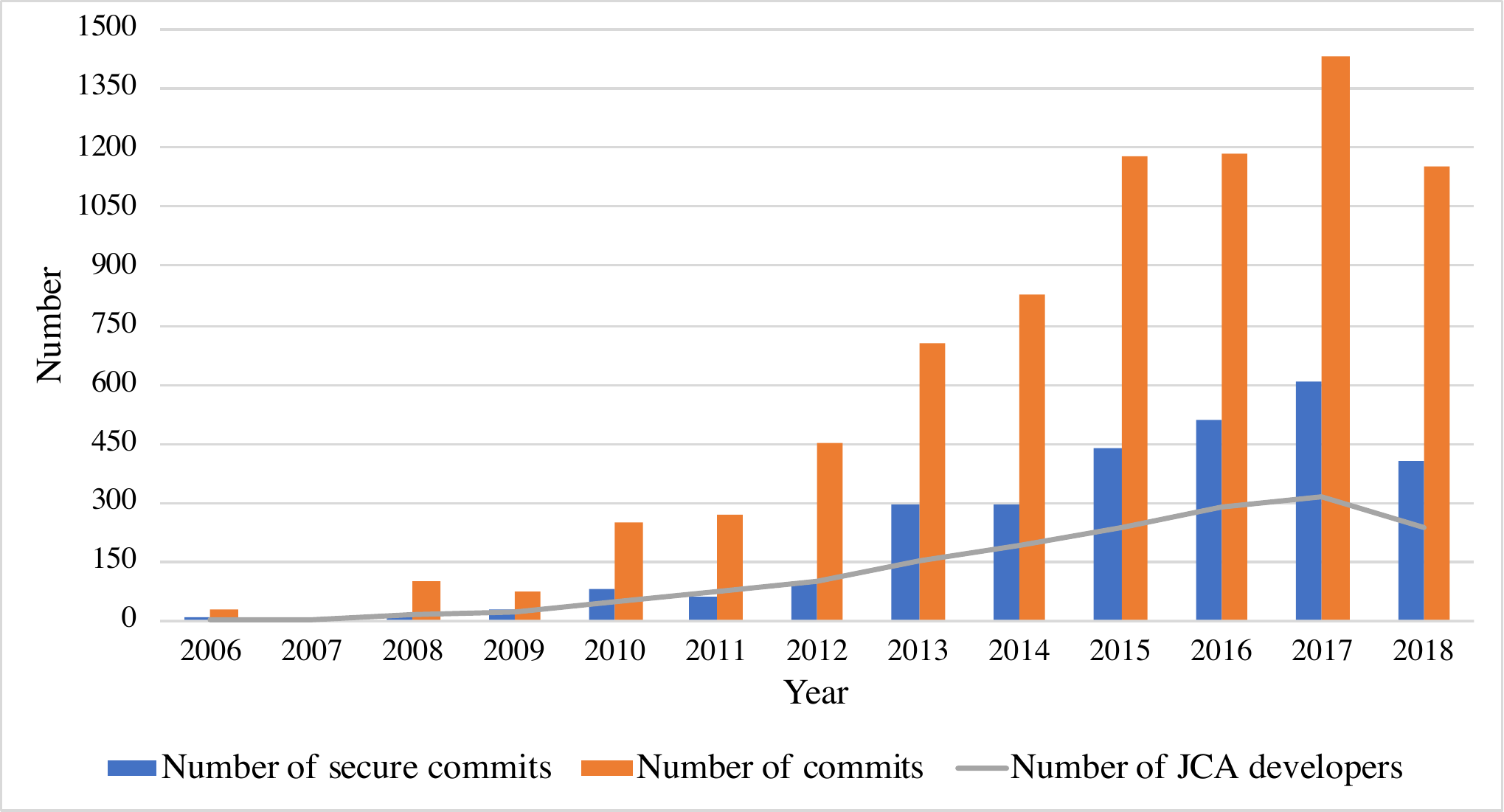}
\caption{The number of secure, total number of commits, and number of JCA developers in each year}\label{fig:annualreport}
\end{figure}

\autoref{fig:annualreport} depicts the number of commits and developers in each year.
As expected, the increase in the number of developers, the number of commits increases as well.
This increase may be due to the emergence of having more cryptographic features in software systems.
However, we assume developers do not necessarily learn how to properly use crypto APIs as the number of secure commits are much lower than the total number of commits in each year.

\autoref{tab:projects} depicts the numbers of secure and buggy projects and commits, and their totals.
We found an average of 1.7 distinct API usages per project with a standard deviation of 1.3.
There is an average number of 3.9 commits per project with a standard deviation of 7.4, of which an average of 1.4 are secure commits and 2.5 are buggy.

\begin{table}[h]
\center
\caption {The status of projects and commits} \label{tab:projects} 
\begin{tabular}{@{}lrrr@{}}
  & Secure & Buggy & Total \\ \midrule
Projects & 642   & 1,682  & 2,324  \\
Commits (LoC)  & 3,263  & 5,897  & 9,160  \\ 
\end{tabular}
\end{table}

\autoref{tab:developers} presents the numbers of distinct developers who always committed secure code, always buggy code, or both secure and buggy code, as well as the total numbers of commits made by each group of JCA developers.
We observed that 27.41\% of developers consistently used the JCA correctly.
The mean value of commits by these developers was 2 with a standard deviation of 1.9.
About 42\% 
of developers had mistakes in every commit.
These developers on average made 3.4 commits with a standard deviation of 3.8. 
The remaining 491 developers had a mean value of 12.13 commits with a standard deviation of 13.67. These developers did not perform well either: the chance of a secure commit is above 45\% only for about half of them (\ie 52.34\%).


\begin{table}[h]
\caption {The status of developers and their commits} \label{tab:developers} 
\center
\begin{tabular}{@{}lrr@{}}
                          &Developer &Commits (LoC)  \\ \midrule
Always secure commits   & 440           & 909           \\
Always buggy commits   & 647           & 2,293       \\
Secure and buggy commits & 491           & 5,958       \\  
\end{tabular}
\end{table}

\begin{figure}
\includegraphics[width=1\linewidth,trim=4 4 4 4,clip]{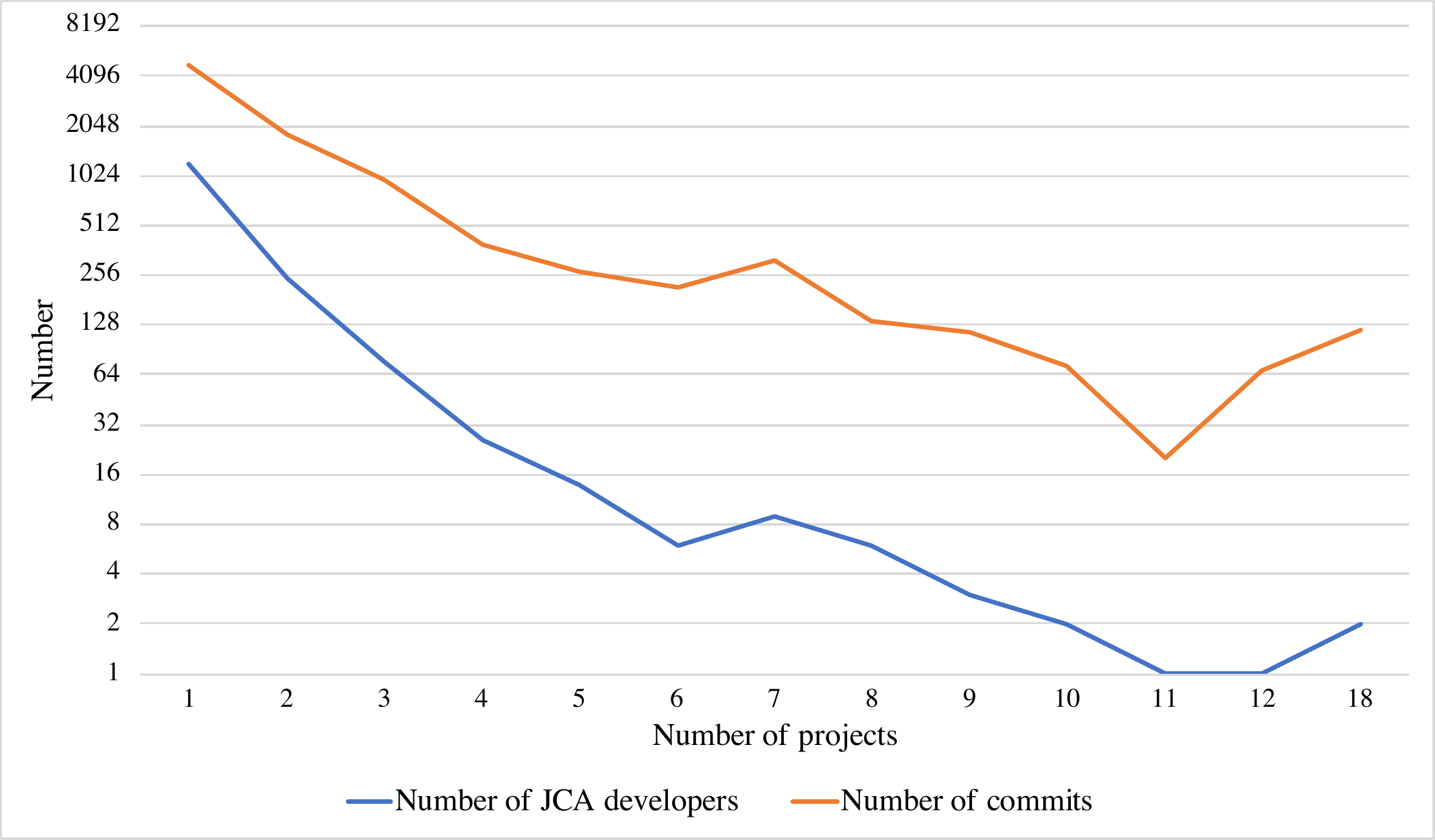}
\caption{The distribution of developers and their commits in different projects}\label{fig:chartfive}
\end{figure}

We also explored how many distinct developers have contributed to different numbers of projects (see \autoref{fig:chartfive}).
We can observe a slight decrease in the number of developers and their total number of commits as the number of involved projects increases.
On average 1.18 JCA developers contributed to each project, and each developer contributes to an average of 1.4 projects.
The highest rate of contributions to JCA projects belongs to two developers who have contributed to 18 projects and produced 118 commits.
In our dataset, 98.13\% of the population of developers and 88.48\% of commits are involved in five or fewer projects.

\begin{figure}
\centering
\includegraphics[width=1\linewidth,trim=4 4 4 4,clip]{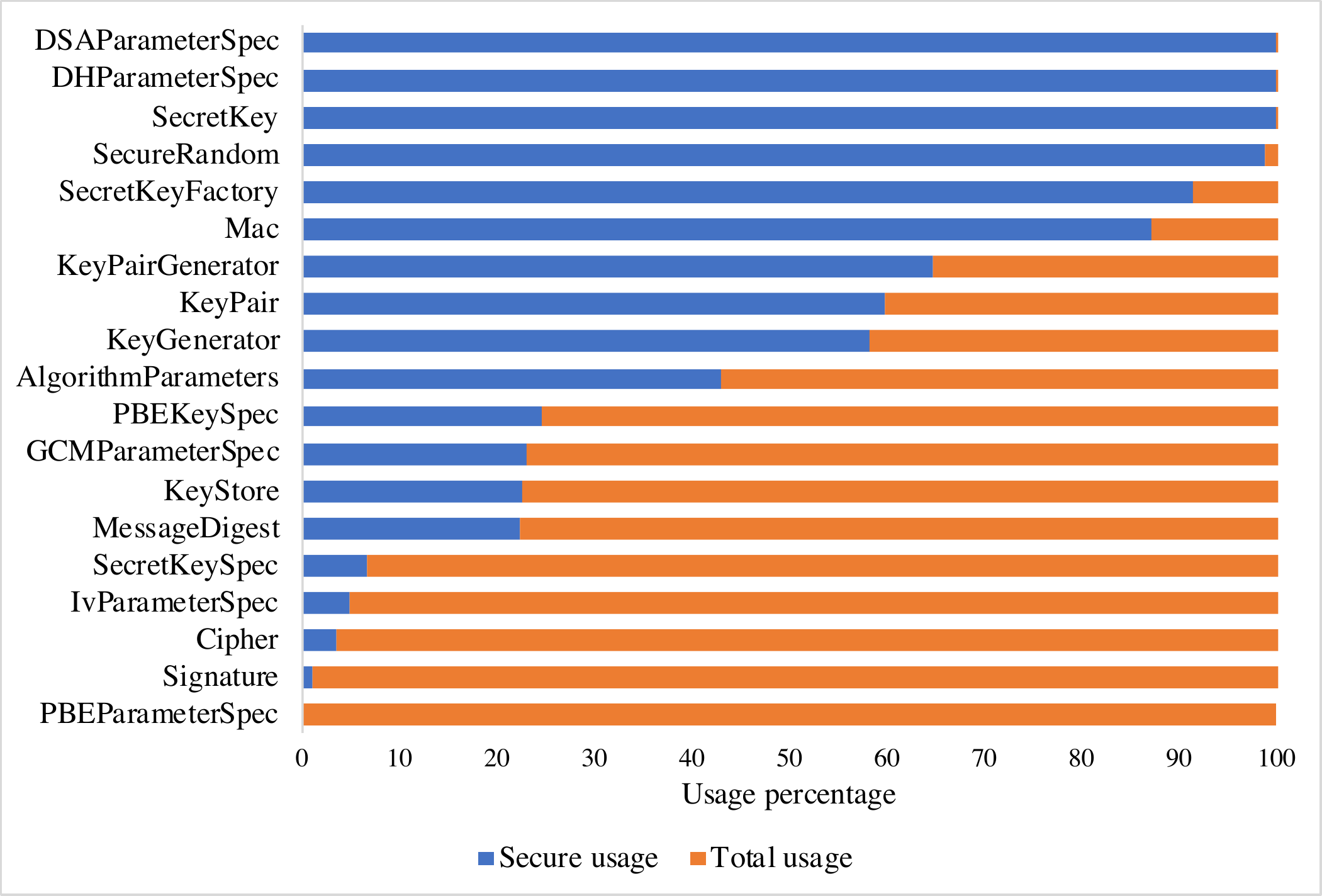}
\caption{The secure versus total number of each API use in percentage}
\label{fig:chartjcaapi2}
\end{figure}

Finally, 
we found that, among the 19 cryptography APIs, \code{MessageDigest} was used the most \ie 2,859 times by 790 developers.
The \code{Cipher}, \code{SecureRandom}, and \code{SecretKeySpec} are the next top APIs that were used more than 1,000 times.
%
\autoref{fig:chartjcaapi2}  presents the distribution of each API use.
We can see that \code{DSAParameterSpec}, \code{DHParameterSpec}, \code{SecretKey}, and \code{SecureRandom}  were almost always used correctly.
Similarly, developers showed promising performance in using the \code{SecretKeyFactory}, and \code{MAC} APIs, \ie at least 87\% usages were correct.
Developers seem to have severe difficulties in using about half of the APIs whose correct usages were less than 25\%.
The correct usages of five APIs namely, \code{SecretKeySpec}, \code{IvParameterSpec}, \code{Cipher}, \code{Signature}, and \code{PBEParameterSpec} were at most 6.58\%.

\boxit{We found that, on average, of 3.9 crypto uses in each project, 2.5 are not secure.
Developers have difficulties in using certain APIs, which require further investigations at API level to understand the reasons underpinning this dilemma.}

\subsection{Factors Influencing Developer Performance}

We define a \emph{performant developer} as one who makes more secure than buggy commits.
Our assumption is that by making more commits, contributing to more projects, working with a wide range of crypto APIs to accomplish different scenarios, and finally being engaged more days with cryptographic APIs can perhaps increase the experience of developers. 
We therefore collected four factors, namely the numbers of (1) JCA commits, (2) APIs used, (3) projects, and (4) days a developer committed, and studied whether these factors, which we assumed they account for developer experience, have an impact on the performance (number of buggy and secure commits) of developers in using cryptography.

We studied the correlation only between each of the four aforementioned factors, and the numbers of secure and buggy commits so as to find factors highly correlated with developer performance.
Both visual inspection and application of the Shapiro-Wilk test confirmed that our data is not normally distributed.
We therefore used the Spearman correlation which does not assume that the data follow a normal distribution \cite{zar2005spearman}.
For every two variables, it generates a number between 1 and $-$1 depending on whether the relationship is positive or negative, respectively.

\begin{table}[]
\caption{The Spearman correlation matrix} \label{tab:spearmancorr} 
\tiny
\begin{tabular}{|l|c|c|c|c|c|c|c|}
\hline
            & \# JCA commits  & \# Project & \# Days & \# API used  & \# Secure  & \# Buggy  \\ \hline 
\# JCA commits       &       & 0.42    & 0.47 & 0.75 & \cellcolor{highlight} 0.53   & \cellcolor{highlight} 0.74        \\ \hline 
\# Project     & 0.42  &         & 0.55 & 0.28 & 0.3    & 0.27        \\ \hline
\# Days   & 0.47  & 0.55    &      & 0.43 & 0.34   & 0.31        \\ \hline 
\# API  used  & 0.75  & 0.28    & 0.43 &      & \cellcolor{highlight} 0.6    & \cellcolor{highlight} 0.51         \\ \hline
\# Secure   & 0.53  & 0.3     & 0.34 & 0.6  &        & -0.051       \\ \hline
\# Buggy  & 0.74  & 0.27    & 0.31 & 0.51 & -0.051 &      \\ \hline
\end{tabular}
\end{table}

\autoref{tab:spearmancorr} depicts the correlation matrix between pairs of variables.
The variables with a correlation score greater than 0.5 or less than $-$0.5 with secure and buggy commits are the number of JCA commits and the number of JCA APIs used.
In the following we study whether these two factors really account for developer performance.


\subsubsection{Number of commits} 

We grouped developers by quartiles of a boxplot which reflects the distribution of developers based on their numbers of commits.
Correspondingly, the numbers of their commits can be grouped into three categories ranging from 2 to 4, 5 to 8, or 9 to more number of commits.

\begin{figure}
\includegraphics[width=1\linewidth,trim=4 4 4 4,clip]{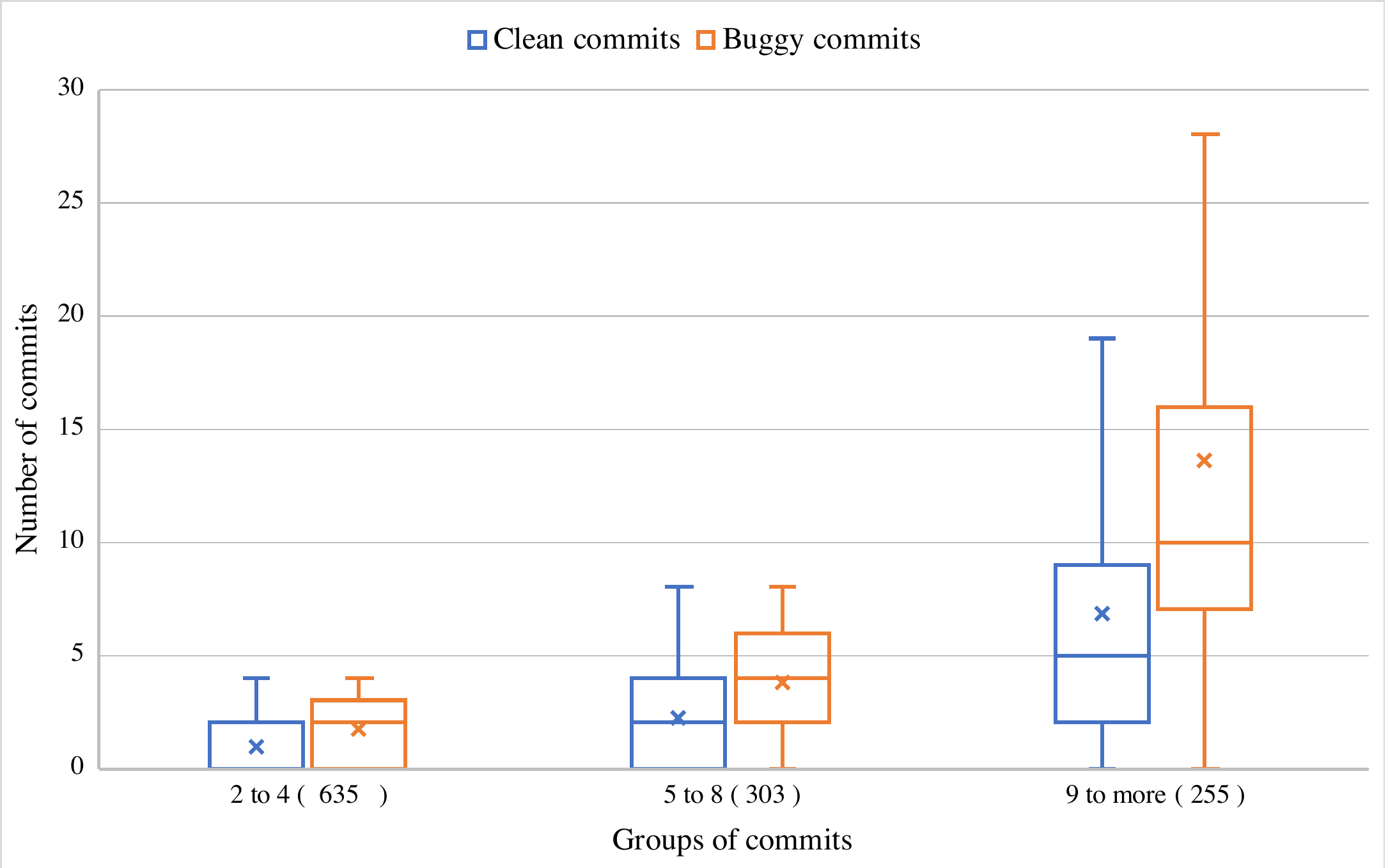}
\caption{Secure and buggy commits based on the number of JCA commits grouping}\label{fig:cleannbuggybox}
\end{figure}

\autoref{fig:cleannbuggybox} presents these groups, and the numbers of secure and buggy commits in each group.
Visually, it is clear that all three groups exhibit an increase in the numbers of secure and buggy commits as the total number of commits increases while the median and the average numbers of buggy commits are always more than those of secure commits.
In the first group, the median of secure commits is zero while it improved in the latter groups.
In the last two groups with more commits, the upper whiskers of secure commits show that the maximum number of secure commits is fewer than the maximum number of buggy commits.

To establish statistical evidence, we pose the following null hypothesis: \emph{the groups of commits are from identical populations}.
To avoid the dispersion of the absolute numbers of secure and buggy commits, we evaluated this hypothesis from the \emph{developer performance} perspective \ie the number of secure commits divided by the total number of commits by each developer.
Therefore, we checked whether the performance of developers in the different groups is the same.

We used a rank-based nonparametric Kruskal-Wallis test since the three groups did not have a normal distribution and this violates one of the assumptions of the one-way ANOVA test.
The test resulted a p-value of 0.001315 (chi-square 13.269, and df 2), which indicates strong evidence that at least one of the groups is different.

\begin{table}[]
\center
\caption {The Wilcoxon signed rank test result for the performance \vs counts} \label{tab:tbltwelve} 
\begin{tabular}{|l|r|r|r|}
\hline
\multicolumn{1}{|c|}{} & First group & Second group  \\ \hline
Second group           & 0.0273     &                    \\ \hline
Third group            & 0.0018     & 1.000                 \\ \hline
\end{tabular}
\end{table}

We  carried out post hoc analysis to find out which pairs of groups are significantly different.
We used pairwise Wilcoxon signed rank comparison with the Bonferroni adjustment for the p-value.
\autoref{tab:tbltwelve} presents the result.
We can verify that the first group has a significant difference compared to the second and third group (the p-values are considerably less than 0.05).
Nevertheless, the second group comes from an identical population as the third group.

\begin{figure}
\includegraphics[width=1\linewidth,trim=4 4 4 4,clip]{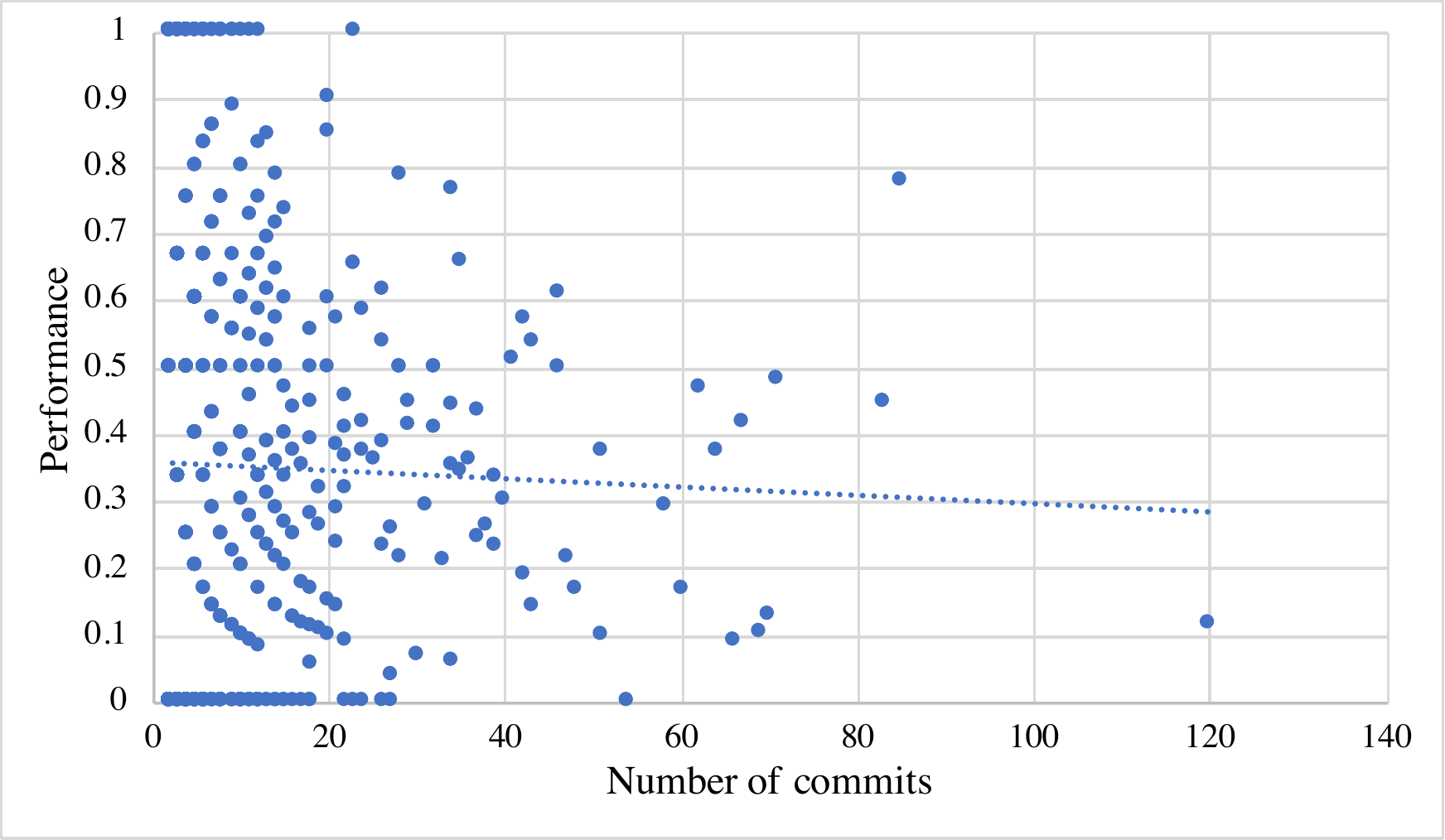}
\caption{Performance \vs number of JCA commits}\label{fig:perfcount}
\end{figure}

\autoref{fig:perfcount} presents the performance of developers versus the number of commits they have made.
In the plot, we can see that many developers with a low number of commits have a performance close to either zero or one.
The reason for this is clearly that, with few commits, every secure or buggy commit has a much greater impact on the performance than for developers with many commits.
For this reason the first group, with only few commits, has a different profile than the others.
The trend line shows that with the increase in the number of commits the performance slightly decreases; we believe that with more commits, more APIs are used, which consequently increases the chance of bugs.
Therefore, if we do not consider the first group, we can accept the null hypothesis, and conclude that the performance of developers in the remaining groups is identical.
In other words, we cannot conclude that developer experience, as measured by number of commits, clearly influences their performance.




\boxit{To summarize, although the number of commits is correlated with the \emph{number} of secure and buggy commits, we did not observe any correlation between the number of commits and developer performance, \ie the \emph{fraction} of secure commits.}


\subsubsection{Number of APIs}  

\begin{figure}
\includegraphics[width=1\linewidth,trim=4 4 4 4,clip]{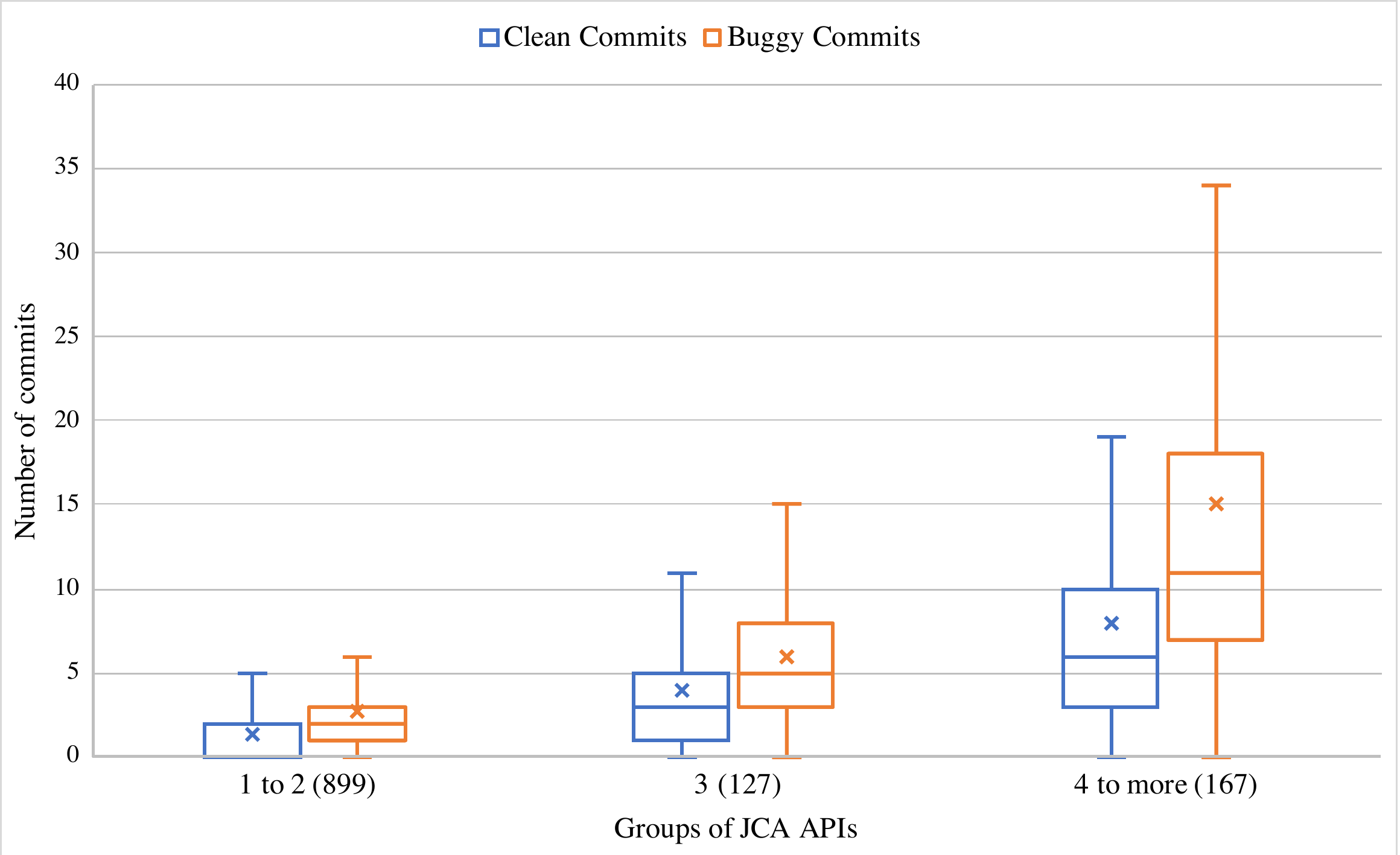}
\caption{Secure and buggy commits based on API grouping}\label{fig:cleannbuggyboxapi}
\end{figure}

We grouped developers by quartiles of a boxplot, which reflects their distribution based on the number of different JCA APIs that they use.
Correspondingly, the numbers of their commits can be grouped into three categories ranging from 1 to 2, 3, or 3 to more APIs.
\autoref{fig:cleannbuggyboxapi} presents these groups, and the numbers of secure and buggy commits in each group.
Similar to the number of commits, the median number of secure commits in the first group is zero and it increases as the number of APIs increases.
The mean value of buggy commits in each group is always higher than secure commits and the maximum value is notably higher than secure commits.
To establish statistical evidence, we computed the performance of each developer (\ie secure commits divided by total number of commits), and used this metric instead of the absolute number of secure and buggy commits.
We define the following null hypothesis:
\emph{The performance of developers is similar in groups that use a different number of APIs.}
The Kruskal-Wallis test shows a significant difference in at least one of the groups.
In particular, the p-value of 2.956e-06 (Chi-square 25.464) is not even close to the cutoff value (\ie 0.05), which shows a significant difference in the three groups.

\begin{table}[]
\center
\caption {The Wilcoxon signed rank test result for performance \vs API} \label{tab:tblfourteen} 
\begin{tabular}{|l|r|r|}
\hline
\multicolumn{1}{|c|}{} & First group & Second group \\ \hline
Second group           & 0.00056     &              \\ \hline
Third group            & 0.00024     & 0.19557      \\ \hline
\end{tabular}
\end{table}

\begin{figure}
\centering
\includegraphics[width=1\linewidth,trim=4 4 4 4,clip]{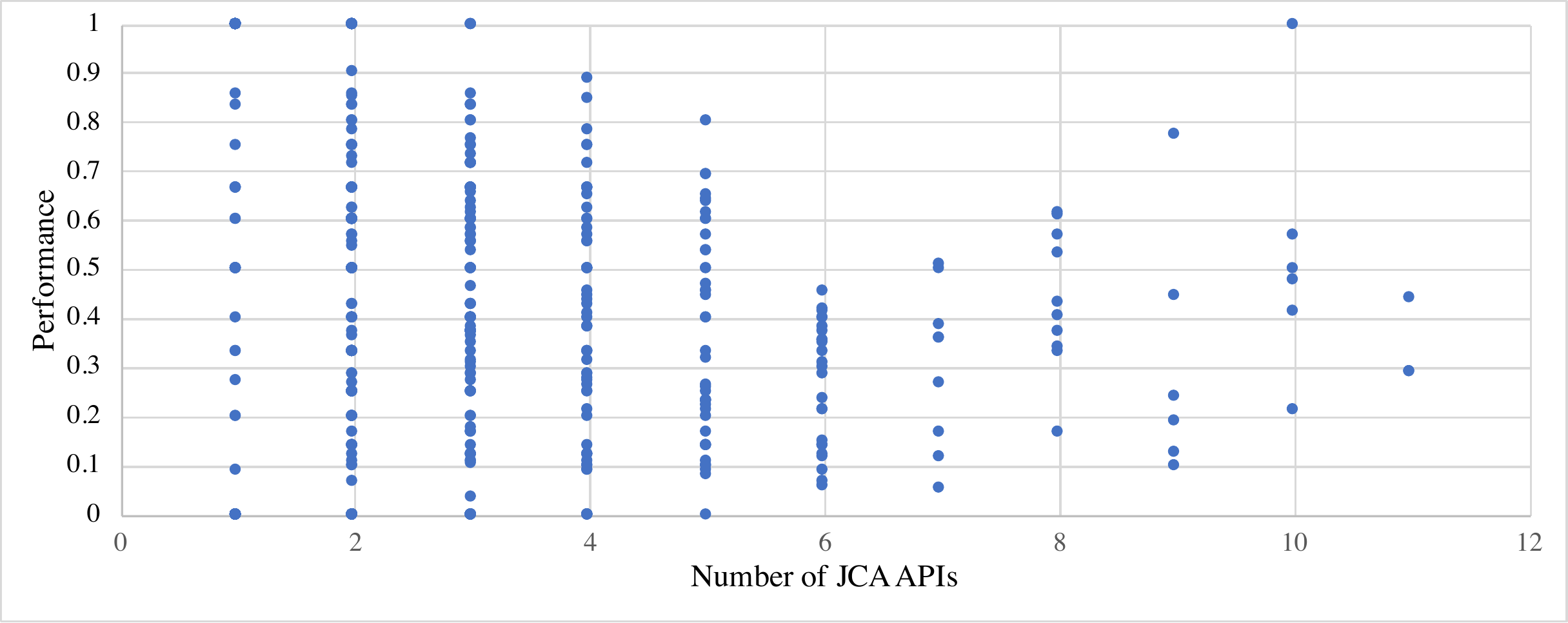}
\caption{Performance \vs number of APIs}\label{fig:perfapi}
\end{figure}

The Wilcoxon signed rank test shows that the first group has a considerable difference with the other groups (See \autoref{tab:tblfourteen}).
Further investigation showed that developers in the first group mostly overlap with those with the lowest number of commits (\ie 2 to 4 commits).
As in the previous subsection, we observe that every commit has a much higher impact on performance for these developers compared to other groups, leading to the difference of the first group with others.
If we account for this phenomenon, we cannot reject the null hypothesis, and conclude that the performance of developers who use different APIs is identical.
\autoref{fig:perfapi} also shows the performance of developers versus the number of APIs that they used.
Clearly, there is no sharp negative or positive trend in the performance of developers can be seen as the number of API increases.

\boxit{To sum up, even though the APIs are strongly correlated with the \emph{number} of secure and buggy commits, we did not observe any correlation between this factor and developer performance}

\subsection{Threats to Validity}

We analyzed 2,324 Java projects that use JCA APIs.
We found these projects through the contributors of some initial projects that were identified in previous work.
We did not define any criteria for selecting the remaining projects, but they may not represent the whole Java ecosystem.
Also, we only focused on Maven projects and did not look for Gradle or ANT projects. 
We relied on \CG for misuse detection as it provides state-of-the art static analyses and covers a wide range of API misuses.
Whereas certain limitations are inherent in static analysis in general, we did not check for the existence of false positives in the results as the authors have done so and found the tool to be fairly precise.
We aborted the analysis of each project after 15 minutes. 
Increase in the timeout may yield more projects to be analyzed.
We used the \textit{git blame} command in order to identify the last developer who has committed a change to that line of code, and did not look into the commit history.
Though the last commit can be due to some refactoring or maintenance purposes, we studied several cases and did not find any other changes concerning crypto APIs.
Moreover, commits can be an outcome of several works by different people locally where the resulting commit is the only visible one.

\section{Related Work}
\label{sec:related}

%
Nadi \etal surveyed 11 developers who asked crypto-related questions on Stack Overflow, as well as 37 developers who used Java's cryptography APIs, and  found that developers are confident in selecting the right cryptography concepts, but they have difficulties in using certain cryptographic algorithms correctly.
They concluded that crypto APIs are generally perceived to be too low-level, and developers prefer more task-based solutions~\cite{Nadi2016}.
Acar \etal manually analyzed the security properties of Stack Overflow posts, which their participants used to resolve pre-defined challenges, and they automatically analyzed  200,000 randomly sampled apps from the Google Play market \cite{acar2016you}.
They observed that real-world Android developers use Stack Overflow and are unwilling to use official Android API documentation.
Gorski \etal conducted a controlled experiment with 53 participants to study the effectiveness of API-integrated security advice.
Their results exhibit that their approach considerably helps 73\% of the participants to produce more secure code by employing security guidance from the documentation \cite{gorski2018developers}.
Yasemin \etal suggested that documentation with secure and easy-to-use code examples of cryptographic APIs could help developers in using the API correctly~\cite{Yasemin2017}.
%
Researchers have also developed static analysis tools that facilitate the use of cryptography.
Rahman \etal developed a static analysis tool named RIGORITYJ to reduce false positives in finding cryptographic misuses \cite{Rahaman2018chiron}.
They evaluated 46 large-scale Apache projects and 6,181 Android apps.
They found 96\% of the Android applications to be vulnerable, and found 1,961 warnings in Apache projects.
Egele \etal developed a static analysis application named CryptoLint, which was tested on 11,748 applications using cryptographic APIs~\cite{Egele}.
They showed that 88\% of the applications have at least one cryptographic API misuse.
Kr\"uger \etal proposed a tool called \CG, an Eclipse plugin that enables developers to securely use cryptographic API \cite{cgcrypt}.
\CG also provides developers with secure code templates.
Furthermore, \CG comes with CrySL, a description language for correct usages of cryptographic APIs, which empowers cryptographic experts to expand correct usages for other APIs\cite{KrugerS0BM18}.
Shuai \etal created a prototype system named Crypto Misuse Analyzer(CMA), which can efficiently identify the crypto misuse vulnerabilities based on the pre-defined model \cite{ShaoDGYS14}.
In their experiment, more than half of the analyzed Android applications have cryptographic misuse vulnerabilities.


\section{Research Plan}
\label{sec:plan}

Additional studies are needed to find a relation between developer experience and developer performance.
First, we need to understand how developers with good performance obtain this level of proficiency to cope with the challenges of using crypto APIs. 
We therefore would like to survey such developers, as well as those with low performance. 
In particular, we would like to learn about the background of developers, the training programs they have participated in, the tools they use, and the information sources that they consult with.
In previous research, API complexity has been studied generally.
We have shown that certain APIs seem to be much more complex for developers to use.
We would like to delve into the reasons that harden such API uses.
In particular, we will investigate whether the difference is in the API design, the quality of documentation, or available code examples.

\section{Conclusions}
\label{sec:conclusion}

We investigated 2,324 open-source Java projects whose code contains usages of Java Cryptography Architecture (JCA).
We discovered that, on average, of 3.9 crypto uses in each project, 2.5 are not secure, and that developers have great difficulties in using more than half of the APIs.
We also studied four factors that rationalize developer experience namely the number of JCA commits, API diversity, the number of projects developers are involved in, and the frequency of committed lines of code.
We found that none of these factors influence developer performance in this domain.

\section*{Acknowledgment}

We gratefully acknowledge the financial support of the 
Swiss National Science Foundation for the project ``Agile Software Assistance'' (SNSF project No.\,200020-181973, Feb.\,1, 2019 - April 30, 2022).


\bibliographystyle{IEEEtran}
\bibliography{thebibliography}

\begin{thebibliography}{10}
\providecommand{\url}[1]{#1}
\csname url@samestyle\endcsname
\providecommand{\newblock}{\relax}
\providecommand{\bibinfo}[2]{#2}
\providecommand{\BIBentrySTDinterwordspacing}{\spaceskip=0pt\relax}
\providecommand{\BIBentryALTinterwordstretchfactor}{4}
\providecommand{\BIBentryALTinterwordspacing}{\spaceskip=\fontdimen2\font plus
\BIBentryALTinterwordstretchfactor\fontdimen3\font minus
  \fontdimen4\font\relax}
\providecommand{\BIBforeignlanguage}[2]{{%
\expandafter\ifx\csname l@#1\endcsname\relax
\typeout{** WARNING: IEEEtran.bst: No hyphenation pattern has been}%
\typeout{** loaded for the language `#1'. Using the pattern for}%
\typeout{** the default language instead.}%
\else
\language=\csname l@#1\endcsname
\fi
#2}}
\providecommand{\BIBdecl}{\relax}
\BIBdecl

\bibitem{Egele}
\BIBentryALTinterwordspacing
M.~Egele, D.~Brumley, Y.~Fratantonio, and C.~Kruegel, ``An empirical study of
  cryptographic misuse in {Android} applications,'' in \emph{Proceedings of the
  2013 ACM SIGSAC Conference on Computer Communications Security}, ser. CCS
  '13.\hskip 1em plus 0.5em minus 0.4em\relax New York, NY, USA: ACM, 2013, pp.
  73--84. [Online]. Available: \url{http://doi.acm.org/10.1145/2508859.2516693}
\BIBentrySTDinterwordspacing

\bibitem{KrugerS0BM18}
S.~Kr{\"{u}}ger, J.~Sp{\"{a}}th, K.~Ali, E.~Bodden, and M.~Mezini, ``{CrySL}:
  An extensible approach to validating the correct usage of cryptographic
  {APIs},'' in \emph{32nd European Conference on Object-Oriented Programming,
  {ECOOP} 2018, July 16-21, 2018, Amsterdam, The Netherlands}, 2018, pp.
  10:1--10:27.

\bibitem{Rahaman2018chiron}
S.~Rahaman, Y.~Xiao, K.~Tian, F.~Shaon, M.~Kantarcioglu, and D.~Yao,
  ``{CHIRON}: Deployment-quality detection of {Java} cryptographic
  vulnerabilities,'' \emph{arXiv preprint arXiv:1806.06881}, 2018.

\bibitem{ShaoDGYS14}
S.~Shao, G.~Dong, T.~Guo, T.~Yang, and C.~Shi, ``Modelling analysis and
  auto-detection of cryptographic misuse in {Android} applications,'' 2014, pp.
  75--80.

\bibitem{Chatzikonstantinou:2016}
A.~Chatzikonstantinou, C.~Ntantogian, G.~Karopoulos, and C.~Xenakis,
  ``Evaluation of cryptography usage in {Android} applications,'' in
  \emph{International Conference on Bio-inspired Information and Communications
  Technologies}, 2016, pp. 83--90.

\bibitem{LazarCWZ14}
D.~Lazar, H.~Chen, X.~Wang, and N.~Zeldovich, ``Why does cryptographic software
  fail?: a case study and open problems,'' 2014, pp. 7:1--7:7.

\bibitem{Nadi2016}
S.~{Nadi}, S.~{Kr\"uger}, M.~{Mezini}, and E.~{Bodden}, ```{Jumping} through
  hoops': Why do {Java} developers struggle with cryptography {APIs}?'' in
  \emph{2016 IEEE/ACM 38th International Conference on Software Engineering
  (ICSE)}, May 2016, pp. 935--946.

\bibitem{Yasemin2017}
Y.~{Acar}, M.~{Backes}, S.~{Fahl}, S.~{Garfinkel}, D.~{Kim}, M.~L. {Mazurek},
  and C.~{Stransky}, ``Comparing the usability of cryptographic {APIs},'' in
  \emph{2017 IEEE Symposium on Security and Privacy (SP)}, May 2017, pp.
  154--171.

\bibitem{zar2005spearman}
J.~H. Zar, ``Spearman rank correlation,'' \emph{Encyclopedia of Biostatistics},
  vol.~7, 2005.

\bibitem{acar2016you}
Y.~Acar, M.~Backes, S.~Fahl, D.~Kim, M.~L. Mazurek, and C.~Stransky, ``You get
  where you're looking for: The impact of information sources on code
  security,'' in \emph{2016 IEEE Symposium on Security and Privacy (SP)}.\hskip
  1em plus 0.5em minus 0.4em\relax IEEE, 2016, pp. 289--305.

\bibitem{gorski2018developers}
P.~L. Gorski, L.~L. Iacono, D.~Wermke, C.~Stransky, S.~M{\"o}ller, Y.~Acar, and
  S.~Fahl, ``Developers deserve security warnings, too: On the effect of
  integrated security advice on cryptographic {API} misuse,'' in
  \emph{Fourteenth Symposium on Usable Privacy and Security ($\{$SOUPS$\}$
  2018)}, 2018, pp. 265--281.

\bibitem{cgcrypt}
\BIBentryALTinterwordspacing
S.~Kr\"{u}ger, S.~Nadi, M.~Reif, K.~Ali, M.~Mezini, E.~Bodden, F.~G\"{o}pfert,
  F.~G\"{u}nther, C.~Weinert, D.~Demmler, and R.~Kamath, ``{CogniCrypt}:
  Supporting developers in using cryptography,'' in \emph{Proceedings of the
  32Nd IEEE/ACM International Conference on Automated Software Engineering},
  ser. ASE 2017.\hskip 1em plus 0.5em minus 0.4em\relax Piscataway, NJ, USA:
  IEEE Press, 2017, pp. 931--936. [Online]. Available:
  \url{http://dl.acm.org/citation.cfm?id=3155562.3155681}
\BIBentrySTDinterwordspacing

\end{thebibliography}

\end{document}